\documentclass[cits]{PoS}

\usepackage{graphicx}
\usepackage{epstopdf}
\title{
\vspace*{-5mm}
\begin{flushright}
{\normalsize\textsc{KEK-TH-1181} \textsc{RBRC-692}
}
\end{flushright}
Nucleon form factors and structure functions with $N_f$=2+1 dynamical domain wall fermions}

\ShortTitle{Nucleon form factors and structure functions with $N_f$=2+1 dynamical DWF}

\author{Takeshi Yamazaki\thanks{Speaker for Nucleon form factors with $N_f$=2+1 domain wall fermions}\\
        University of Connecticut,
        Physics Department,
        Storrs, Connecticut 06269-3046, USA\\
        \email{yamazaki@phys.uconn.edu}}

\author{Shigemi Ohta%
\thanks{Speaker for Nucleon structure functions with $N_f=$2+1 dynamical domain-wall fermions}\\
Institute of Particle and Nuclear Studies, KEK, Tsukuba, Ibaraki 305-0801, Japan\\
RIKEN-BNL Research Center, Brookhaven National Laboratory, Upton, NY 11973\\
Physics Department, Sokendai Graduate U.\ Adv.\ Studies, Hayama, Kanagawa 240-0193, Japan\\
\email{shigemi.ohta@kek.jp}}

\author{RBC and UKQCD Collaborations} 

\abstract{We report isovector form factors and low moments of structure functions of nucleon in
  numerical lattice quantum chromodynamics (QCD) from the on-going calculations by the
  RIKEN-BNL-Columbia (RBC) and UKQCD Collaborations with (2+1) dynamical flavors of domain-wall
  fermion (DWF) quarks.
We calculate the matrix elements with four light quark masses, corresponding to pion mass values of
  $m_\pi = 330$--$670$ MeV, while the dynamical strange mass is fixed 
at a value close to physical, on (2.7 fm)$^3$ spatial volume.
We found that our axial charge, \(g_A\),  at the lightest mass exhibits a large deviation from the heavier mass results.
This deviation seems to be a finite-size effect as the \(g_A\) value 
scales with a single parameter, $m_\pi L$, the product of pion mass 
and linear spatial lattice size.
The scaling is also seen in earlier 2-flavor dynamical DWF and Wilson quark calculations.
Without this lightest point, the three heavier mass results show only very mild mass dependence and linearly extrapolate to \(g_A=1.16(6)\).
We determined the four form factors, the vector (Dirac), induced tensor (Pauli), axial vector and induced pseudoscalar, at a few finite momentum transfer values as well.
At the physical pion mass the form-factors root mean square radii determined from the momentum-transfer dependence 
are 20--30\% smaller than the corresonding experiments.
The ratio of the isovector quark momentum to helicity fractions, \(\langle x\rangle_{u-d}/\langle x\rangle_{\Delta u - \Delta d}\) is in agreement with experiment without much mass dependence including the lightest point.
We obtain an estimate, 0.81(2), by a constant fit. 
Although the individual momentum and helicity fractions are yet to be renormalized, they show encouraging trend toward experiment.
}

\FullConference{The XXV International Symposium on Lattice Field Theory\\
		 July 30 - August 4 2007\\
		 Regensburg, Germany}

\begin{document}

$\phantom{\speaker{T.~Yamazaki and S.~Ohta}}$

\section{Introduction}

We report isovector form factors and low moments of structure functions of nucleon in numerical lattice quantum chromodynamics (QCD) from the on-going calculations by the RIKEN-BNL-Columbia (RBC) and UKQCD Collaborations \cite{Antonio:2006px,Allton:2007hx,Antonio:2007tr} with degenerate up and down and a heavier strange flavors represented by domain-wall fermions (DWF) \cite{Kaplan:1992bt,Shamir:1993zy,Furman:1995ky}.

The isovector form factors are defined in the following two equations:
\begin{eqnarray}
\langle p | V_\mu (0) | p \rangle &=&
\overline{u}_p \left[ \gamma_\mu F_1(q^2) + \sigma_{\mu\nu} q_\nu
F_2(q^2) / 2m_N \right] u_p,
\label{eq:vector}\\
\langle p | A_\mu (0) | p \rangle &=&
\overline{u}_p \left[ \gamma_\mu \gamma_5 G_A(q^2) + i q_\mu \gamma_5
G_P(q^2) \right] u_p,
\label{eq:axial_vector}
\end{eqnarray}
where $V_\mu = \overline{u} \gamma_\mu u - \overline{d} \gamma_\mu d$
and $A_\mu = \overline{u} \gamma_\mu \gamma_5 u - 
\overline{d} \gamma_\mu \gamma_5 d$ are isovector vector and axial vector currents, respectively.
These form factors are experimentally measured in neutron decays and other electroweak transitions of nucleon \cite{PDBook}.
All the form factors can be calculated numerically on the lattice \cite{Sasaki:2003jh,Sasaki:2007gw}.

The structure functions are measured in deep inelastic lepton scatterings off nucleon \cite{PDBook}.
For their definitions we refer the readers to an earlier RBC publication \cite{Orginos:2005uy} and references cited there in.
In this report we discuss some of their isovector low moments such as the momentum fraction \(\langle x\rangle_{u-d}\), helicity fraction \(\langle x\rangle_{\Delta u - \Delta d}\), transversity \(\langle 1\rangle_{\delta u - \delta d}\) and twist-3 \(d_1\).

\section{Formulation}

The matrix elements corresponding to linear combinations of 
the form factors eqs.(\ref{eq:vector}) and (\ref{eq:axial_vector}) 
are determined from the ratio of 
the three-point and two-point functions~\cite{Hagler:2003jd}
\begin{equation}
R^{\mathcal{O,P}}_\mu (t,{\bf q}) = 
\frac{ G^{\mathcal{O,P}}_\mu(t,{\bf q}) }{ G_2^G(t^\prime-t_0,{\bf 0})}
\left[
\frac{ 
G_2^L(t^\prime-t,{\bf q}) G_2^G(t-t_0,{\bf 0}) G_2^L(t^\prime-t_0,{\bf 0}) }{
G_2^L(t^\prime-t,{\bf 0}) G_2^G(t-t_0,{\bf q}) G_2^L(t^\prime-t_0,{\bf q}) }
\right]^{\frac{1}{2}},
\end{equation}
where $G^{\mathcal{O,P}}_\mu(t,{\bf q})$ is three-point function,
\begin{equation}
G^{\mathcal{O,P}}_\mu(t,{\bf q}) = 
\frac{1}{4}\mathrm{Tr}\left[
\mathcal{P}\langle 0 | \chi(t^\prime,{\bf 0})
\mathcal{O}(t,{\bf q}) \overline{\chi}(t_0,-{\bf q})
\rangle
\right],
\end{equation}
with the current $\mathcal{O}=V_\mu,A_\mu$, the projector 
$\mathcal{P} = \frac{1+\gamma_t}{2}(\mathcal{P}_t), 
\frac{1+\gamma_t}{2}\gamma_5\gamma_z(\mathcal{P}_{5z})$,
the spatial momentum transfer ${\bf q}$,
and $\chi$ being the nucleon field.
$t_0$ and $t^\prime$ are the sources of the three-point function,
and $G_2^{L,G}$ is the two-point function with the point($L$) or
the gauge invariant Gaussian smearing($G$) operator sink.
All the sources for the two- and three-point functions 
are calculated with the gauge invariant Gaussian smearing operator.
$R_\mu^{\mathcal{O,P}}(t)$ at $t_0 \ll t \ll t^\prime$ will be constant,
and corresponds to the linear combination of the form factors
multiplied by kinematic factors.
The form factor is obtained by solving the linear equation.
The details of the equations are in Ref.~\cite{Sasaki:2007gw}.

The axial charge is calculated by the ratio of the three-point function 
$G^{A,\mathcal{P}_{5z}}_z(t)/G^{V,\mathcal{P}_{t}}_t(t) = Z_V g_A / Z_A$
at zero momentum transfer.
This ratio gives the renomalized axial charge $g_A$, because the axialvector and vector currents share a common renormalization, $Z_A = Z_V$, up to second order discretization error thanks to the well-preserved chiral symmetry of the domain wall fermions.
We confirm that these renormalization constants agree
within 0.5\% accuracy in the chiral limit.

The matrix elements related to the structure functions,
such as the momentum fraction, 
helicity distribution, moment of transversity and $d_1$,
are calculated by the ratio of the three-point function
to the two-point function, $G^{\mathcal{O^{\prime},P}}(t)/G_2^G(t^\prime)$,
at zero momentum transfer.
Definitions of the operators $\mathcal{O^\prime}$ are listed in detail 
in Ref.~\cite{Orginos:2005uy}.
At the lightest quark mass $m_f = 0.005$ 
the three-point function with the temporal direction
of the conserved vector current~\cite{Blum:2001sr} $\mathcal{V}_t$ at
${\bf q}=0$ is used for the 
denominator of the ratio instead of $G_2^G(t^\prime)$.
This ratio gives same matrix element because a relation,
$G^{\mathcal{V,P}_t}_t(t)/G_2^G(t^\prime) = 1$, is satisfied for  $t_0 \ll t \ll t^\prime$.

\section{Numerical setup}

The calculations are performed on the QCDOC dedicated computers \cite{QCDSPOC} at RIKEN-BNL Research Center (RBRC) and University of Edinburgh.
Descriptions of the ensembles used are found in RBC-UKQCD publications \cite{Antonio:2006px,Allton:2007hx,Antonio:2007tr}.
We use a combination of Iwasaki rectangular gauge action \cite{Iwasaki:1983ck} with the gauge coupling set at 2.13 and domain-wall fermion \cite{Kaplan:1992bt,Shamir:1993zy,Furman:1995ky} quarks with the domain-wall height set at 1.8.
A \(24^3\times 64\) lattice is used.
We fix the dynamical strange mass at 0.04 and generated four different ensembles with degenerate up and down mass each at 0.03, 0.02, 0.01 and 0.005 in lattice units.
From these we estimate the lattice cut off to be about \(a^{-1}\) of 1.73(3) GeV~\cite{Lin:2007pos}
with the $\Omega^-$ baryon mass.
The \(24^3\) lattice spatial volume thus corresponds to a physical volume of about  \((2.74(4) {\rm fm})^3\). 
The residual quark mass that parameterizes the mixing of the two domain walls across the \(L_s=16\) fifth dimension is estimated to be about 0.0031.
The physical strange mass is estimated as about 0.035(1) plus the 0.0031 residual mass from the squared mass ratio of kaon and \(\Omega^-\).

\begin{figure}
\begin{center}
\includegraphics*[width=.5\textwidth]{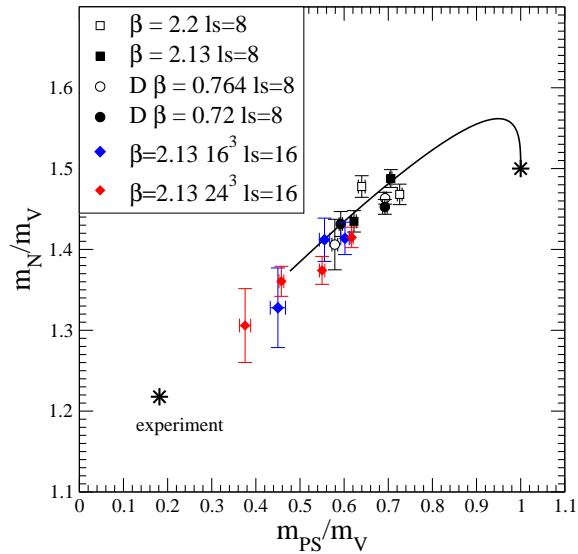}
\vspace{-7mm}
\end{center}
\caption{An Edinburgh plot obtained from the present and related RBC and UKQCD joint ensembles with (2+1)-flavor dynamical DWF quarks.}
\label{fig:eplot}
\end{figure}
For the nucleon matrix element calculation reported here, 106, 98, 356 and 178 configurations are used respectively for light quark mass values  of 0.03, 0.02, 0.01 and 0.005.
We see a reasonable behavior in both pion and nucleon mass in their approach to the chiral limit, as shown in Figure \ref{fig:eplot}.
We obtained pion masses of \(m_\pi\) = 0.67, 0.56, 0.42 and 0.33 GeV,  and nucleon \(m_N\) = 1.56, 1.39, 1.25 and 1.15 GeV, respectively from the ensembles with the quoted light quark mass values.

Four measurements are carried out for each configuration to improve
statistics, with the source set at time slices $t_0= 0, 16, 32, 48$ 
or $t_0 = 8, 19, 40, 51$ except at $m_f = 0.005$.
In order to reduce computational cost at $m_f = 0.005$, we employ 
non-relativistic quark field and a double source method 
where the two sources, either  at
$(0, 32)$ or $(16, 48)$, are set for one quark propagator.
In the three-point function the source and sink operators are separated 
by 12 time slices to reduce excited-state contamination as much as possible.

\section{Form factor results}

\subsection{Axial charge}
\label{sec:g_A}

\begin{figure}[b]
\begin{center}
\scalebox{0.30}[0.3]{
\includegraphics*{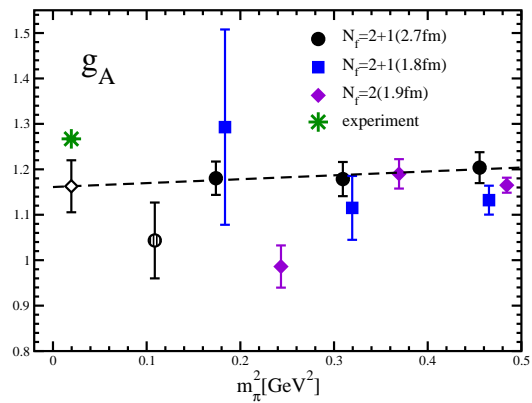}}
\vspace{-7mm}
\end{center}
\caption{Axial charges with 2+1 flavor and 2 flavor.
Dashed line represents linear chiral extrapolation of
larger volume, 2+1 flavor data without lightest point.
\label{fig:gagv_mpi}}
\end{figure}

Figure~\ref{fig:gagv_mpi} shows our result of the axial charge normalized by
$Z_V$.
The (2.7 fm)$^3$ volume data are well determined 
and the statistical uncertainties are less than 8\%.
The data are almost independent of 
the pion mass squared 
except the lightest point.
The lightest pion mass data is about 15\% smaller
than the other pion mass data. 
An earlier 2-flavor calculation by the RBC Collaboration \cite{Lin:2007} 
with the spatial volume (1.9 fm)$^3$ and $1/a=1.7$ GeV
showed a similar trend, but with the downward behavior setting in at
a heavier pion mass than the current 2+1 flavor case.

We suspect that this pion mass dependence driving the axial charge away from
the experiment at light quark mass values is caused by the finite lattice volume:
In general such finite volume effect is expected to grow as we set the quark mass lighter as such lighter quarks fluctuate more.
This interpretation is not inconsistent with the observed behavior of the (2+1)- (present) and 2-flavor (\cite{Lin:2007}) DWF results.
Furthermore the finite volume effect is larger on smaller spatial volume 
when the quark mass is same.
More quantitatively, we observe in the figure
that the 2-flavor result from the (1.9 fm)$^3$ volume
significantly decreases at $m_\pi = 0.24$ GeV$^2$, while the (2+1)-flavor
results from the (2.7 fm)$^3$ volume does not even at $m_\pi^2 = 0.17$ GeV$^2$.
The similar behavior was also observed in an earlier small-volume,
quenched study~\cite{Sasaki:2003jh}.

Also shown in the figure is a set of (2+1)-flavor results from a  smaller volume~\cite{nuc_paper}, (1.8 fm)$^3$, with the pion mass, spatial volume and lattice spacing comparable to the 2-flavor calculations.
However, the result at the lightest pion mass suffers from a large statistical fluctuation and prevents us from deciding whether there is a similar finite-size effect here.

\begin{figure}[b]
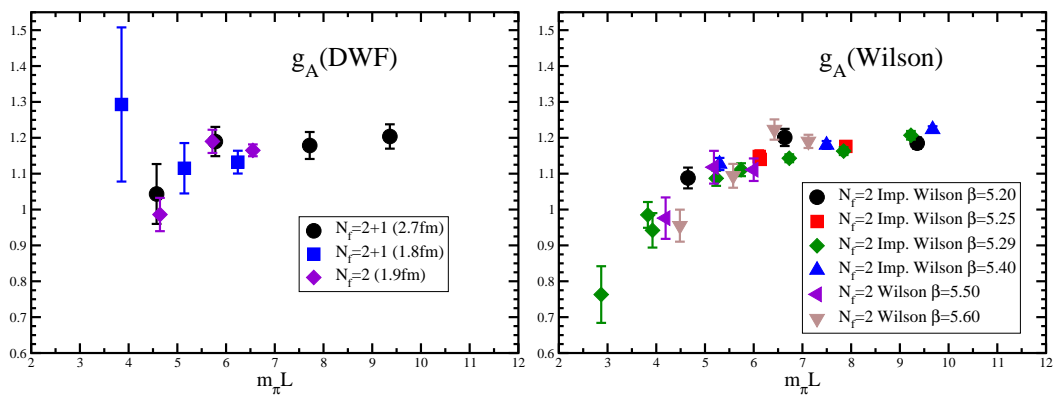

\begin{center}
\scalebox{0.30}[0.3]{
\includegraphics*{Fig/mpiL_D.eps}}
\scalebox{0.30}[0.3]{
\includegraphics*{Fig/mpiL_W.eps}}
\vspace{-7mm}
\end{center}
\caption{Axial charges with our dynamical domain wall(left panel) and
dynamical Wilson fermions(right panel) obtained
by LHPC/SESAM~\cite{Dolgov:2002zm} 
and QCDSF~\cite{Khan:2006de} collaborations
as a function of $m_\pi L$.
\label{fig:gagv_mpiL}}
\end{figure}

In order to compare the results from  the (2+1)- and 2-flavor calculations, we plot the axial charge against a dimensionless quantity, $m_\pi L$, as presented in the left panel of Figure~\ref{fig:gagv_mpiL}.
The (2+1)-flavor, larger volume results and the 2-flavor ones align well with each other, and suggest a monotonic dependence on \(m_\pi L\): in other words, an $m_\pi L$ scaling.
The (2+1)-flavor results from the smaller volume are also consistent with this $m_\pi L$ scaling except for the lightest point that suffers from large statistical fluctuation.
This large statistical fluctuation itself can be another manifestation of a large finite-size effect.
Nevertheless we plan to improve the statistics of this lightest point so as to test the reliability of this $m_\pi L$-scaling interpretation.

We also plot the axial charge calculated with 2 flavors of dynamical Wilson and improved Wilson quarks respectively by LHPC/SESAM~\cite{Dolgov:2002zm} and QCDSF~\cite{Khan:2006de} 
collaborations against $m_\pi L$ (see the right panel in Figure \ref{fig:gagv_mpiL}.)
These calculations were performed at various different spatial volumes, 
pion masses, and the gauge couplings.
Like our (2+1)- and 2-flavor DWF results in the above, all of these were calculated at unitary points where the sea and valence quark masses are equal, $\kappa_{sea} = \kappa_{val}$.
These Wilson quark results also seem to suggest a similar scaling in $m_\pi L$, with a downward behavior setting in at $m_\pi L$ around and below 6.

In the above we discussed that the downward shift away from the experiment of the axial charge at lighter quark mass values may well be caused by finite lattice volumes.
We found the axial charge is well described by a monotonic scaling in the dimensionless parameter, $m_\pi L$.
This $m_\pi L$ scaling seems common among both DWF and Wilson dynamical quark calculations performed at unitary points.
Therefore, we seem to have a strong case to suspect a large finite-size effect in nucleon electroweak matrix elements in the quark mass range relevant for extrapolating to the physical or chiral point.

Thus we carry out chiral extrapolation of the axial charge using only the larger volume results and without the lightest pion mass point.
We simply use a linear function of the pion mass squared, because there now are only three available data points which do not suggest any non-linear behavior in the pion mass squared.
The extrapolation is presented also in Figure~\ref{fig:gagv_mpi}:
we obtain $g_A = 1.16(6)$ 
at the physical pion mass $m_\pi = 0.14$ GeV.
This extrapolated value is 8.3\% smaller than the experiment.
Note this deviation may also be caused by a finite volume effect which may be present even at the heavier quark mass values used in the estimation.

We need more detailed study to clarify this possibly large finite volume effect, not only in the axial charge but also in other form factors and structure functions.
We plan a larger volume calculation at the same cut off in the near future.

\subsection{Isovector Dirac form factor}

\begin{figure}[b]
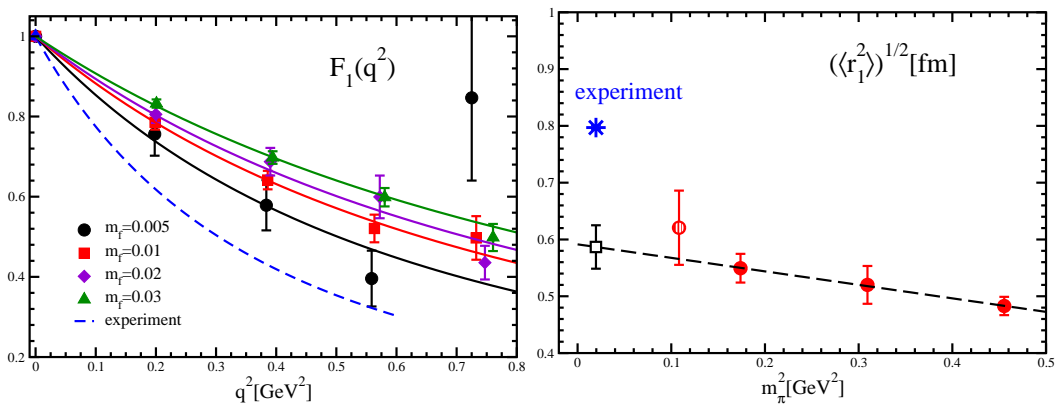

\begin{center}
\scalebox{0.30}[0.3]{
\includegraphics*{Fig/F_v_f.eps}}
\scalebox{0.30}[0.3]{
\includegraphics*{Fig/mfr_1.eps}}
\vspace{-7mm}
\end{center}
\caption{
Isovector Dirac form factor and Dirac rms radius
are presented in left and right panels, respectively.
Dashed line in right panel is chiral extrapolation
without lightest pion mass data.
\label{fig:F_v}}
\end{figure}

The left panel of Figure \ref{fig:F_v} shows our result of the isovector Dirac form factor at the four quark mass values plotted against the momentum transfer squared, $q^2$.
The form factor is renormalized by $Z_V$, in other words, normalized by $1/F_1(0)$.
Traditionally the experimental Dirac form factor is considered to be approximated well by a dipole form,
\begin{equation}
F_1(q^2) = 1/(1 + q^2/M_1^2)^2,
\end{equation}
where $M_1$ denotes the dipole mass.
This traditional experimental dipole fit is shown in the figure also, represented by the dashed line with $M_1 = 0.857(8)$ GeV~\cite{PDBook}.
Thus it is convenient to fit the present, lattice-calculated Dirac form factor by the dipole form as well. The fit results are presented in the figure as solid lines for each quark mass value.
With a mild dependence on the quark mass, there is tendency for the calculated results to approach the experimental line as the quark mass is decreased.

The Dirac root mean square (rms) radius is determined by the dipole mass by a relation:
$\sqrt{ \langle r^2_1 \rangle} = \sqrt{12}/M_1$. 
Thus the experimental value is  0.794(4) fm.
The right panel of Figure \ref{fig:F_v} shows our results for the Dirac rms radius obtained by the dipole fit.
The pion mass dependence is again mild as it was in the dipole fit discussed in the above.
While the result at the lightest pion mass is trending toward the 
experiment with a large statistical error, we cannot exclude that large
finite-size effect as discussed in the above subsection~\ref{sec:g_A}
does not affect the data.
We nevertheless decided to exclude this point from our chiral 
extrapolation which is represented by the dashed line in the figure: 
we employ a linear chiral extrapolation
due to the mild $m_\pi^2$ dependence of the heavier data, and 
obtain $\sqrt{ \langle r^2_1 \rangle} = 0.59(4)$ 
fm.
The result reproduces about 74\% of the experimental value at the physical pion mass.
As can be seen from the figure, the lightest point is consistent with this fit with its large statistical fluctuation.

\subsection{Isovector Pauli form factor}

\begin{figure}[b]
\begin{center}
\scalebox{0.30}[0.3]{
\includegraphics*{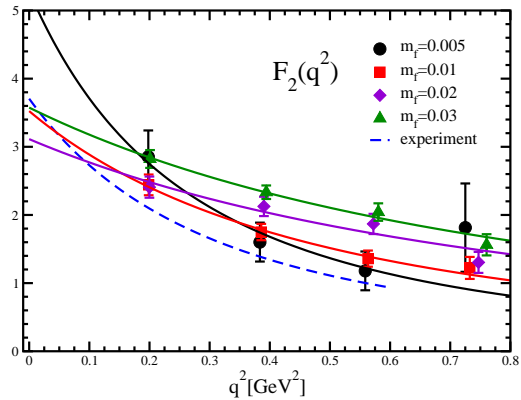}}
\vspace{-7mm}
\end{center}
\caption{
Isovector Pauli form factor as a function of momentum
transfer squared.
\label{fig:F_t}}
\end{figure}

The isovector Pauli form factor is the induced tensor part of the 
vector current matrix element.
We can calculate it at the same time with the Dirac form factor except 
at zero momentum transfer where the kinematics prevent us.
Figure~\ref{fig:F_t} presents the results of our calculation of this form factor at each
quark mass plotted against the four momentum transfer squared.
The form factor is renormalized by $Z_V$ again.

The results, unlike the Dirac form factor, suffer from statistical fluctuation.
Yet they can be fitted by the dipole form,
\begin{equation}
F_2(q^2) = F_2(0) / ( 1 + q^2 / M_2^2 )^2.
\end{equation}
The fit results are shown in the figure.
There is a large quark mass dependence:
the form factor decreases as the quark mass is decreased.
The three heavier quark mass results seem to approach the experiment, shown as the dashed line with
the dipole mass 
$M_2 = 0.78(2)$ 
GeV~\cite{PDBook}.
The lightest quark mass result is an exception, however.
Note it suffers large fluctuation.


%
\begin{figure}[t]
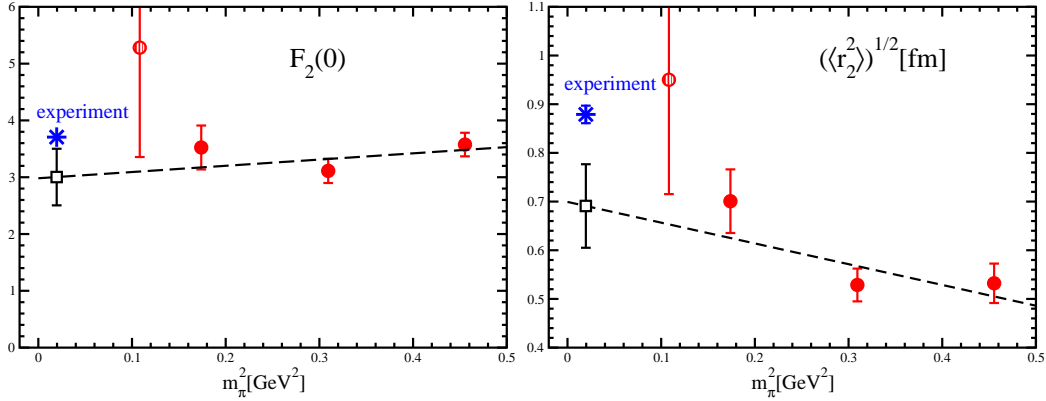

\begin{center}
\scalebox{0.30}[0.3]{
\includegraphics*{Fig/mfF_t.eps}}
\scalebox{0.30}[0.3]{
\includegraphics*{Fig/mfr_2.eps}}
\vspace{-7mm}
\end{center}
\caption{
Pauli form factor at zero momentum transfer and 
Pauli rms radius are presented in left and right panels,
respectively.
Dashed lines are chiral extrapolations
without lightest pion mass data.
\label{fig:F_t_fit}}
\end{figure}

We determine the anomalous magnetic moment $F_2(0)$ and the Pauli rms radius $\sqrt{ \langle r^2_2 \rangle} = \sqrt{12}/M_2$ from the dipole fit.
Figure~\ref{fig:F_t_fit} shows the former in the left and the latter in the right panel as functions of the pion mass squared.

The anomalous magnetic moment shows only a mild mass dependence, and is almost consistent with the experiment, $F_2(0) = \mu_p - \mu_n - 1 = 3.70589$ even at the heaviest point ($\mu_p$ and $\mu_n$ are the magnetic moments of proton and neutron respectively.)
Though we again cannot exclude the possibility of large finite-size effect at the lightest point, the result there is almost consistent with the linear fit to the rest which gives an extrapolation
to the physical point, $F_2(0) = 3.0(5)$.

In contrast, the rms radius shows a strong mass dependence approaching the experiment as the pion mass decreases.
Again we cannot exclude the possibility of large finite-size effect at the lightest point.
Omitting the lightest point the linear fit gives an extrapolated rms radius of $\sqrt{ \langle r^2_2 \rangle} = 0.69(9)$ 
fm at the physical point.

These extrapolated values of anomalous magnetic moment and Pauli rms radius are about 20~\% lower than the corresponding experiments.  While the former almost catches the experiment within one standard deviation, the latter is about two standard deviations away.

\subsection{Isovector axial vector form factor}

\begin{figure}[b]
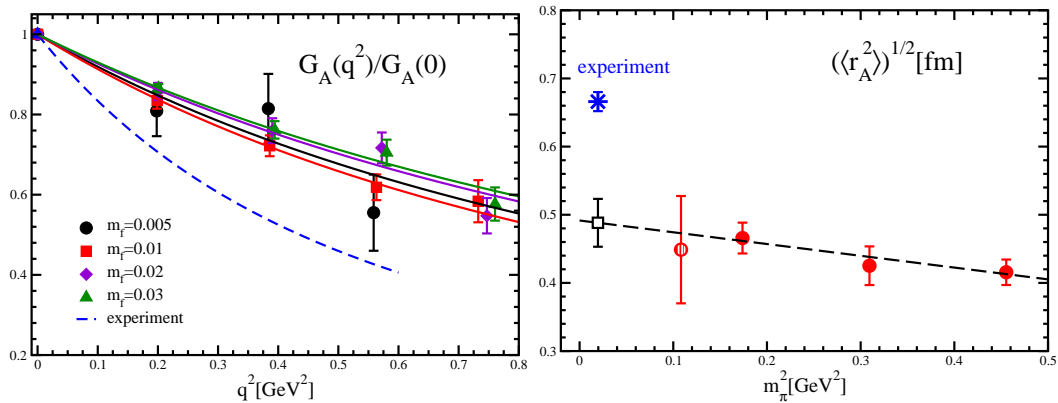

\begin{center}
\scalebox{0.30}[0.3]{
\includegraphics*{Fig/NF_a_f.eps}}
\scalebox{0.30}[0.3]{
\includegraphics*{Fig/mfr_a.eps}}
\vspace{-7mm}
\end{center}
\caption{
Isovector axial vector form factor and axial vector rms radius
are presented in left and right panels,
respectively.
Dashed line in right panel is chiral extrapolation
without lightest pion mass data.
\label{fig:F_a}}
\end{figure}

In this section we focus only on the momentum transfer dependence
of the axial vector form factor:
We normalize the form factor by its value at zero momentum transfer respectively for each quark mass.
The left panel of Figure \ref{fig:F_a} shows the results after these normalizations, $G_A(q^2)/G_A(0)$.
The experimental form factor is again traditionally considered to be fitted well by the dipole form,
\begin{equation}
G_A(q^2)/G_A(0) = 1/(1+q^2/M_A^2)^2,
\label{eq:dipole_ga_ff}
\end{equation}
with the experiments giving $M_A = 1.03(2)$
GeV~\cite{Bernard:2001rs} for the axial vector dipole mass.
The experimental fit is shown by the dashed line in the figure.

Fits to the calculations with the dipole form are represented by the solid lines and well describe the calculations for the heavier three quark mass values.
The lightest quark mass results suffer large statistical errors and fluctuations.
This is because the presented values are normalized by \(G_A(0)\), a quantity directly proportional to the axial charge which itself suffers large statistical errors and fluctuations as discussed in the subsection \ref{sec:g_A}. 
The mass dependence here is even milder than in the case of the vector current Dirac form factor.

The axial charge rms radius is determined from the dipole mass,
$\sqrt{ \langle r^2_A \rangle} = \sqrt{12}/M_A$,
and is 0.666(14) fm in the experiment.
The calculated axial charge rms radii from the fits are shown in the right
panel of Figure \ref{fig:F_a} plotted against the pion mass squared.
While the result increases as the pion mass decreases,
it is about 30\% smaller than the experiment.
The lightest pion mass data is omitted in the following chiral extrapolation,
because we cannot rule out a large systematic error stemming from the suspected large finite-size effect.
However the result would not change, as can be seen from the figure, if we included the point.
We carry out a linear fit and extrapolation with the heavier three mass values and obtain 0.49(4) 
fm at the physical pion mass.
The result reproduces 73\% of the experiment.

\subsection{Isovector induced pseudoscalar form factor}

\begin{figure}[b]
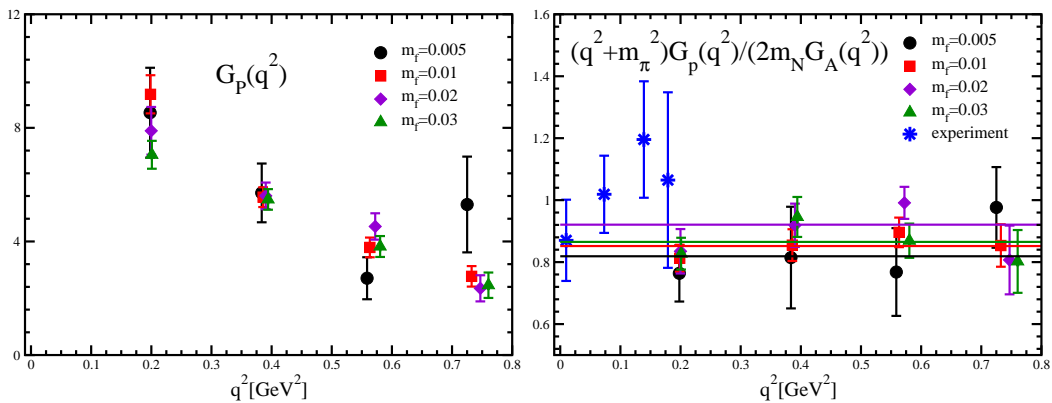

\begin{center}
\scalebox{0.30}[0.3]{
\includegraphics*{Fig/F_p.eps}}
\scalebox{0.30}[0.3]{
\includegraphics*{Fig/NF_p.eps}}
\vspace{-7mm}
\end{center}
\caption{
Isovector induced pseudoscalar form factor 
and ratio of pseudoscalar form factor
to axial vector form factor are presented in left and right panels,
respectively.
\label{fig:F_p}}
\end{figure}

The induced pseudoscalar form factor, $G_P(q^2)$, is obtained as a part of the matrix element
of the axial vector current.
This form factor is expected to have a pion pole, so its momentum-transfer dependence should be different from the other form factors.

The left panel of Figure \ref{fig:F_p} shows the calculated $G_P(q^2)$ renormalized with $Z_V$
as plotted against the momentum transfer squared at each quark mass.
Note that the values at the smallest $q^2$ are almost 8 and much larger than other form factors.
In addition, while it may be hard to observe due to large errors, the values at the smallest $q^2$ increase as the quark mass is decreased except at the lightest quark mass. 
This behavior is consistent with pion pole dominance.

The induced pseudoscalar form factor is related to the axial vector form factor through the so-called partially conserved axial vector current (PCAC) relation which is a manifestation of spontaneously broken chiral symmetry.
In the traditional PCAC current algebra, the celebrated generalized Goldberger-Treiman relation,
\begin{equation}
G_P(q^2) = 2 m_N G_A(q^2) / ( q^2 + m_\pi^2 ),
\label{eq:pion_pole}
\end{equation}
is obtained at low $q^2$.  
The denominator on the right-hand side of this relation corresponds to the pion pole.
We investigate the relation in our results through a quantity, $( q^2 + m_\pi^2 ) G_P(q^2) / ( 2 M_N G_A(q^2) )$.
If the relation holds we obtain unity for this quantity.
The right panel of Figure \ref{fig:F_p} shows the quantity in our calculation stays close to unity, and has no significant $q^2$
dependence.
We can simply fit these results by a constant for each quark mass respectively, whose results are presented in the figure as well.
All the fit results are consistent with the experiments~\cite{Andreev:2007wg,Czarnecki:2007th,Choi:1993vt} within the larger error of the experiments.

\subsection{$\pi NN$ coupling} 
\begin{figure}[t]
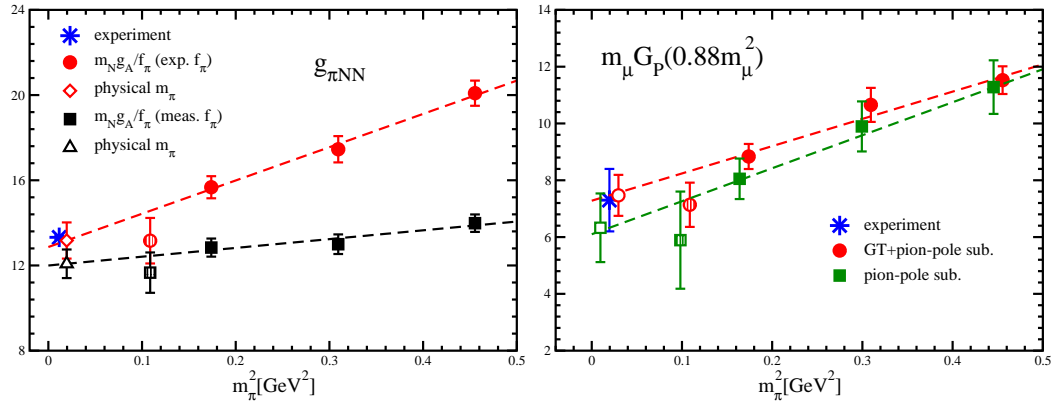

\begin{center}
\scalebox{0.30}[0.3]{
\includegraphics*{Fig/mfgAg_f.eps}}
\scalebox{0.30}[0.3]{
\includegraphics*{Fig/mfg_p.eps}}
\vspace{-7mm}
\end{center}
\caption{
Left panel is the $\pi NN$ coupling evaluated by GT relation with
experimental and measured $f_\pi$.
Right panel is induced pseudoscalar coupling for muon capture determined with
generalized GT relation and induced pseudoscalar form factor.
Dashed lines are chiral extrapolations
without lightest pion mass data.
\label{fig:GT_g_p}}
\end{figure}

The original Goldberger-Treiman (GT) relation \cite{Goldberger:1958tr},
\begin{equation}
g_{\pi NN} f_\pi = 2 m_N g_A,
\end{equation}
equates a combination of quantities at the pion pole, the \(\pi NN\) coupling, $g_{\pi NN}$, and the pion decay constant, $f_\pi$, on the left with another combination of quantities at almost zero momentum transfer, the nucleon mass and axial charge.
As such it suffers from the mismatch in momentum transfer if we substitute the experimental values for the quantities, such as $f_\pi=93$ MeV, \(m_N=940\) MeV and \(g_A=1.269\), to obtain a value for the \(\pi NN\) coupling, $g_{\pi NN}$.
Thus it becomes interesting how much better or worse our lattice calculation does in this regard.

In the left panel of Figure \ref{fig:GT_g_p} we show two such calculations each for the \(\pi NN\) coupling,  $g_{\pi NN}$: one uses the experimental value of $f_\pi$ and another the lattice-calculated values at each quark mass, plotted against the pion mass squared.
The results with the experimental $f_\pi$ exhibit a significant slope in terms of $m_\pi^2$, while that with the calculated $f_\pi$ is almost flat.
In both methods the results at the lightest mass show significant downward shift away from the trend set by the three heavier mass values.
This of course is another manifestation of the large deviation observed in the axial charge which was discussed in detail in subsection \ref{sec:g_A}.
For the chiral extrapolation we simply employ a linear fit form and exclude the lightest mass point.
We obtain the results at the physical pion mass, 
$g_{\pi NN} = 13.2(9)$ 
with the experimental $f_\pi$ and
$g_{\pi NN} = 12.1(7)$ 
with the calculated $f_\pi$.
The extrapolated results at the physical pion mass
are consistent with the experiment obtained from forward $\pi$-$N$
scattering data~\cite{Ericson:2000md}.

\subsection{Muon capture}

The induced pseudoscalar coupling, $g_P = m_\mu G_P(q_c^2)$, is defined with the muon mass $m_\mu$ and the induced pseudoscalar form factor $G_p$ at the momentum transfer squared, $q_c^2 = 0.88 m_\mu^2$ GeV$^2$.
This is for convenience in application for muon capture, $p + \mu^- \to n + \nu_\mu$, where the two-body nature of the process defines the momentum transfer, \(q_c\).

$G_P(q^2)$ can be described by the pion pole dominance form
through the generalized Goldberger-Treiman relation 
(\ref{eq:pion_pole}), and a constant $\alpha$ which corresponds to
the difference from unity of the quantity, $( q^2 + m_\pi^2 ) G_P(q^2) / ( 2 M_N G_A(q^2) )$, at each quark mass.
These were summarized in Figure~\ref{fig:F_p}.
Thus, $g_P$ is determined by
\begin{equation}
g_P = \alpha m_\mu \frac{ 2 m_N G_A(q_c^2) }{ q^2_c + m_\pi^2 }
= \alpha m_\mu \frac{ 2 m_N }{ q^2_c + m_\pi^2 } 
\frac{ g_A }{ ( 1 + q^2_c / M_A^2 )^2 },
\end{equation}
where we use the dipole form of the axial vector form factor 
(\ref{eq:dipole_ga_ff}).
In order to subtract the strong pion mass dependence stemming from
the pion pole, we use the physical pion mass in the pion pole.
On the other hand we use the calculated values for $m_N$, $g_A$, $M_A$ and $\alpha$.
The right panel of Figure \ref{fig:GT_g_p} shows 
that the result denoted by circle is almost linear as a function of the pion mass squared and decreases toward the experiment for the three heavier mass values.
Again the lightest mass result is an exception:
We suspect the significant drop here at the lightest point is caused
by the finite volume effect in $g_A$ as discussed in Sec.~\ref{sec:g_A}.

We also determine $g_P$ from $G_P(q^2)$ directly with the subtraction of the pion pole.
At each quark mass the quantity $G_P(q^2) ( q^2 + m_\pi^2 )$ is fitted by the dipole form, and then extrapolated to $q^2_c$.
In the figure the extrapolated result to $q^2_c$ normalized by the pion pole with the physical pion mass is presented by square symbol.
The result from $G_P(q^2)$ has larger error, whilst the two results agree within the statistical errors at each pion mass.
We carry out chiral extrapolations with a simple linear function of the pion mass squared without the lightest mass.
The results at the physical pion mass are consistent with each other, and also with the recent experiment \cite{Andreev:2007wg} and analysis\cite{Czarnecki:2007th}.

\section{Low moments of the structure functions}

\subsection{Quark momentum and helicity fractions}

\begin{figure}[t]
\begin{center}
\includegraphics*[width=.45\textwidth]{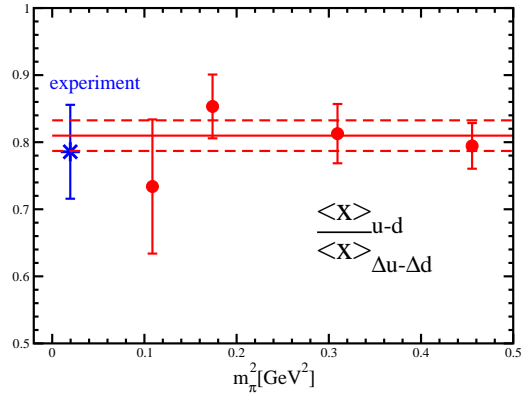}
\vspace{-7mm}
\end{center}
\caption{The naturally renormalized ratio of the isovector quark momentum fraction \(\langle x\rangle_{u-d}\) to helicity fraction \(\langle x\rangle_{\Delta u -\Delta d}\).}
\label{fig:xqdq}
\end{figure}

Let us first discuss the naturally renormalized ratio of the isovector quark momentum fraction \(\langle x\rangle_{u-d}\) to helicity fraction \(\langle x\rangle_{\Delta u -\Delta d}\).
Since the two fractions are related with each other by a chiral rotation, they share a common renormalization.
And because the DWF quarks preserve the chiral symmetry very well, as parameterized by our small residual mass of 0.0031, the ratio is naturally renormalized on the lattice.
Our results are summarized in Figure \ref{fig:xqdq}.
The ratio does not show any appreciable dependence on the quark mass, albeit with large statistical errors, and is in agreement with the experimental value.
The constant fit result with all four pion mass data, 
$\langle x\rangle_{u-d}/\langle x\rangle_{\Delta u -\Delta d} = 0.81(2)$, 
is also consistent with the experiment.

\begin{figure}[b]
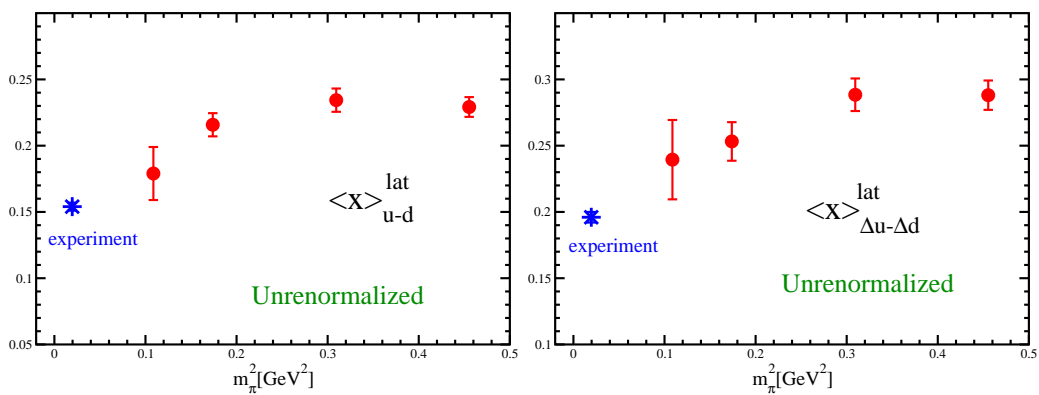

\begin{center}
\includegraphics*[width=.45\textwidth]{Fig/mfxq.eps}
\includegraphics*[width=.45\textwidth]{Fig/mfxdq.eps}
\vspace{-7mm}
\end{center}
\caption{The isovector quark momentum fraction \(\langle x\rangle_{u-d}\) (left) and helicity fraction \(\langle x\rangle_{\Delta u -\Delta d}\) (right), both yet to be renormalized.}
\label{fig:absfractions}
\end{figure}

Each of the fractions itself, unlike the ratio, is yet to be renormalized, but suggests interesting behavior as shown in Figure \ref{fig:absfractions}.
They both trend down toward the experimental value at the lightest quark mass, after staying almost constant for the three heavier mass values.
This may well be related to the finite-size effect we suspect for the form factor values, but not necessarily so as the structure functions probe different, deep inelastic, physics from the elastic form factors.
We will be better able to discuss these quantities after finishing the lattice non-perturbative renormalizations for them in the near future.

\begin{figure}[t]
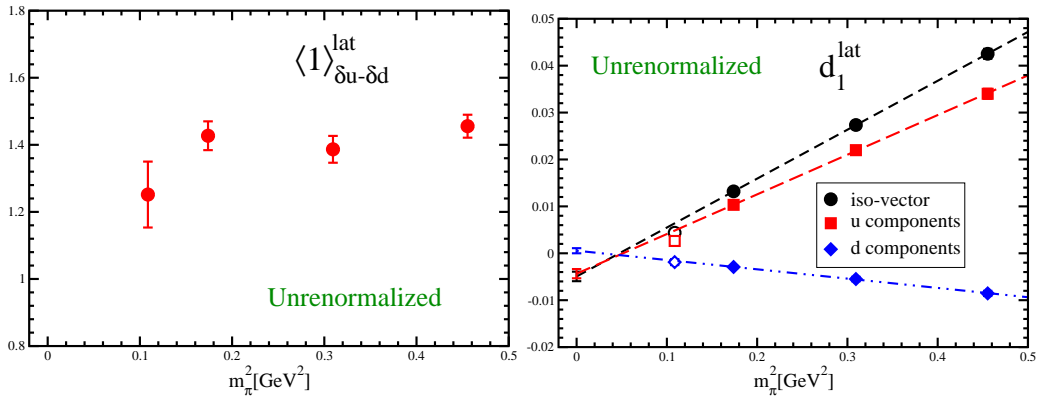

\begin{center}
\includegraphics*[width=.45\textwidth]{Fig/mf1_q.eps}
\includegraphics*[width=.45\textwidth]{Fig/mfd_1.eps}
\vspace{-7mm}
\end{center}
\caption{Transversity, \(\langle 1\rangle_{\delta u - \delta d}\), (left) and twist-3, \(d_1\), (right) moments, unrenromalized.}
\label{fig:transd1}
\end{figure}

\subsection{Transversity}

The transversity, \(\langle 1\rangle_{\delta u - \delta d}\), can be measured by the RHIC Spin experiment in the near future.
In the present calculation it shows a similar behavior with the quark momentum and helicity
fractions: it stays almost constant for the three heavier quark mass values and then trends down at
the lightest mass (see the left panel of Figure \ref{fig:transd1}.)
This quantity is also yet to be renormalized, but will soon be, and then will provide a prediction for the experiments planned in the near future \cite{Matthias}.

\subsection{Twist-3 \(d_1\) moment}

The \(d_1\) moment of the twist-3 part of the polarized structure function, according to Wandzura
and Wilczek \cite{Wandzura:1977qf}, is small when it is calculated perturbatively.
It need not be small under a non-perturbative, confining environment within the nucleon.
Our calculation for this \(d_1\) moment is summarized in the right panel of Figure \ref{fig:transd1}.
The result suggests, though again not renormalized yet, that it is small, consistent with Wandzura and Wilczek.
We note also that the results at the lightest quark mass show some deviation from the linear extrapolations from the three heavier mass values.

\section{Conclusions}

We calculated the isovector form factors and low moments of structure functions of the nucleon with
$N_f=2+1$ dynamical domain wall fermions at 1.7 GeV cutoff on a $(2.7$ fm$)^3$ spatial volume.
The axial charge at the lightest quark mass is about 15\% smaller than the other mass data, and it seems affected by a finite volume effect.
It scales with a single parameter, \(m_\pi L\), the product of pion mass and the linear spatial lattice size.
We confirmed similar scaling in earlier $N_f=2$ dynamical DWF and Wilson fermion calculations.
Without the lightest point the axial charge is estimated as 1.16(6) by a linear extrapolation to the physical point.
The root mean square radii for the form factors except the induced pseudoscalar are determined and
at the physical pion mass are 20--30\% smaller than experiments.
The $\pi NN$ coupling and induced pseudoscalar coupling are found to be consistent with experiments.

We found the renormalized ratio of the isovector quark momentum fraction \(\langle x\rangle_{u-d}\) to helicity fraction \(\langle x\rangle_{\Delta u -\Delta d}\) is in agreement with experiment.
Their individual values, though yet to be renormalized, show an encouraging trend toward the experimental values at the lightest quark mass.
Their non-perturbative renormalization will be completed soon. 
We calculated the bare transversity which will provide a prediction when its non-perturbative renormalization is completed in the near future.
We found the twist-3 \(d_1\) moment is small, consistent with the Wandzura-Wilczek relation.


\section*{Acknowledgments}
We thank the members of the RBC and UKQCD collaborations, present and past, especially T.~Blum, H.~W.~Lin, S.~Sasaki and J.~Zanotti.
We also thank RIKEN, Brookhaven National Laboratory,  University of Edinburgh, UK PPARC and the U.S. DOE for providing the facilities essential for conducting this research.
T.Y. is supported by US DOE grant DE-FG02-92ER40716 and the University of Connecticut.

\providecommand{\href}[2]{#2}\begingroup\raggedright


\end{document}